\documentclass{emulateapj}   

\begin{document}

\title{The Santa Fe Light Cone Simulation Project: I. Confusion and the WHIM in Upcoming Sunyaev-Zel'dovich Effect Surveys}
\author{Eric J. Hallman\altaffilmark{1,2}, Brian W. O'Shea\altaffilmark{3}, 
Jack O. Burns\altaffilmark{2}, Michael L. Norman\altaffilmark{4},
Robert Harkness\altaffilmark{5} \& Rick Wagner\altaffilmark{4}}

\altaffiltext{1}{National Science Foundation Astronomy and
  Astrophysics Postdoctoral Fellow}

\altaffiltext{2}{Center for Astrophysics and Space Astronomy,
  Department of Astrophysics and Planetary Sciences, 
  University of Colorado at Boulder, Boulder, CO 80309; hallman, burns@casa.colorado.edu}

\altaffiltext{3}{Theoretical Astrophysics (T-6), Los Alamos National
Laboratory, Los Alamos, NM 87545; bwoshea@lanl.gov}

\altaffiltext{4}{Center for Astrophysics and Space Sciences,
University of California at San Diego, La Jolla, CA 
92093; mnorman, rpwagner@cosmos.ucsd.edu}

\altaffiltext{5}{San Diego Supercomputing Center, MC0505, 9500 Gilman
  Drive, La Jolla, CA 92093; harkness@sdsc.edu}

\begin{abstract}
 We present the first results from a new generation of simulated large sky coverage ($\sim$100 square degrees) Sunyaev-Zeldovich effect (SZE) cluster surveys
using the cosmological adaptive mesh refinement N-body/hydro code Enzo.
We have simulated a
very large (512$^3$h$^{-3}$Mpc$^3$) volume with unprecedented dynamic
range. We have generated simulated light cones to match the resolution and
sensitivity of current and future SZE instruments. Unlike many
previous studies of this type, our
simulation includes unbound gas, where an appreciable fraction of the
baryons in the universe reside.

We have found that cluster line-of-sight overlap may be a
significant issue in upcoming single-dish SZE surveys. Smaller
beam surveys ($\sim$1$\arcmin$) have more than one massive
cluster within a beam diameter 5-10\% of the time, and a larger beam
experiment like Planck has multiple clusters per beam 60\% of the time. 
We explore the contribution of unresolved halos and unbound gas to the
SZE signature at the maximum decrement. We find that there
is a contribution from gas outside clusters of $\sim$16\% per object on average for upcoming surveys. This adds both bias and scatter to the deduced value
of the integrated SZE, adding difficulty in
accurately calibrating a cluster Y-M relationship. 

Finally, we find that in images where objects with M $>$ 5
$\times$10$^{13}$ M$_{\odot}$ have had their SZE signatures removed,
roughly a third of the total SZE flux still remains. This gas exists at least partially in the Warm Hot
Intergalactic Medium (WHIM), and will possibly be detectable with the
upcoming generation of SZE surveys.
\end{abstract}


\section{Introduction}\label{sec:Intro}
Clusters of galaxies form from the highest peaks in the
primordial spectrum of density perturbations generated by inflation in
the early universe. They are the most massive virialized structures in
the universe, and as such are rare objects. 
The number density of galaxy clusters as a function of mass and
redshift is strongly dependent on a number of cosmological
parameters. In particular, counting the abundance of clusters above
some lower mass limit as a function of cluster redshift places
constraints on $\Omega_b$, $\Omega_M$, $\sigma_8$, and the dark energy
equation of state parameter, $w$ \citep{wang, haiman01}.  

Observational measurement of the cluster abundance over large
sky areas and redshift range is required to
generate cosmological parameter constraints which are complementary
to constraints from the cosmic microwave background (CMB), type Ia
supernovae, Big Bang nucleosynthesis (BBN), the Lyman-$\alpha$ forest,
and galaxy redshift surveys \citep{tozzi}. Cluster survey yields
depend on the value of the minimum cluster luminosity probed as a function of
redshift, the scaling between cluster luminosity and mass, the growth function 
of structure, and the redshift evolution
of the comoving volume element, all of which depend on a complex combination
of cosmological parameters \citep{rosati} and intracluster medium
(ICM) physics \citep{ev04}. 

\subsection{Sunyaev-Zel'dovich Effect Surveys}
The Sunyaev-Zel'dovich Effect \citep{sz} is a process by
which the hot electrons trapped in the large dark matter potential
wells of clusters inverse Compton scatter CMB photons to higher
energy, resulting in a low frequency ($<$ 218 GHz) decrement, and a corresponding
high frequency ($>$ 218 GHz) increment in the CMB on the angular scale of the
cluster. The strength of the SZE decrement/increment is characterized
by the Compton $y$ parameter, which results from the line of sight
integral of the thermal pressure
\begin{equation}
y = \int \sigma_T n_e \frac{k_b T}{m_e c^2} dl,
\end{equation}
where $n_e$ is the electron density and $T$ is the gas electron
temperature. We also define the integrated Compton $y$ parameter as
\begin{equation}
Y = \int y dA
\end{equation}
which is the integration over the area subtended by a circle
corresponding to some relevant cluster physical radius. 
The observed temperature fluctuation corresponding to a
give value of  $y$ in a given frequency band is
\begin{equation}
\frac{\Delta T}{T} = y g(x),
\end{equation}
where 
\begin{equation}
g(x) = \left(x \frac{e^x + 1}{e^x -1} - 4\right) [1 + \delta_{sze}(x,T_e)],
\end{equation}
$x = h\nu/kT_{cmb}$ and $\delta_{SZE}$ is a relativistic correction as
described in \citet{itoh}. For our purposes in this paper we have
neglected the relativistic correction, but will explore it in future
work. This correction is small (less than 1\% at the maximum decrement
frequency) for clusters with T $<$ 10 keV. 

The SZE is particularly useful in cosmological studies due to its
near redshift independence \citep{reph,birk,carlstrom}.  Therefore, observations of clusters are
not limited to low redshift as in the X-ray, but can extend to as high
as z $\sim$ 2, where the number of massive clusters becomes small. An additional consequence of the redshift independence
of SZE surveys is that a flux-limited survey is also approximately a mass
limited survey \citep{reph,haiman01}. These two properties make SZE surveys uniquely
valuable for cluster abundance counts and determination of the cluster
mass function with redshift, provided one can obtain independent
optical redshifts for the galaxies in the identified objects. The near redshift
independence of the SZE creates unique complications for large area surveys
which do not seriously affect other types of surveys (e.g., optical and X-ray).  In
particular, the contribution to the sky signal from both low mass and
distant halos, as well as unbound gas, may be a significant source of
confusion.  

There are several
upcoming millimeter wavelength cluster surveys with new telescopes including
the Atacama Pathfinder Experiment Sunyaev-Zeldovich survey (APEX-SZ) \citep{apex}, the South Pole Telescope
(SPT) \citep{spt}, and the Atacama Cosmology
Telescope (ACT) \citep{act}, in addition to the space-based \textit{Planck} Surveyor \citep{planck}, which will conduct large blind surveys of clusters
using the SZE. Table \ref{sz_inst} shows the values
of the relevant instrumental characteristics for the single-dish
survey telescopes we have taken from the literature and used in the
following analysis. For this study, we limit ourselves to results
using a single band, $\sim$144 GHz, where the SZE decrement is
maximal, though these results will generalize to other bands and
to multiwavelength studies, since the SZE from all gas will have roughly the same
spectral signature (modified slightly by relativistic effects). However, for removing contaminating signals (e.g. radio
point sources, CMB) multiwavelength coverage will be very desirable.

The variation in survey characteristics for these instruments has
important consequences for cluster surveys. For example, the distribution of sources detected as
a function of cluster redshift should be different for each
survey. This is because although the SZE surface brightness does not diminish with distance,
the angular size of the objects does vary with redshift, and may be
larger or smaller than the instrument beam for any given cluster. This selection effect is
modified also by the volume sampled as a function of redshift in a
fixed angular field and the growth rate of structure in an $\Lambda$CDM
universe. It is important to understand these selection effects
in order to constrain cosmology. We must be able to determine the
correct distribution of clusters as a function of redshift from the
surveys or systematic errors in estimated cosmological parameters will
result. 

\begin{table}
\caption{Characteristics of Upcoming SZE Surveys}
\begin{center}
\begin{tabular}{cccc}
\hline
\hline
Survey & Angular Coverage & Beam Size ($\sim$144 GHz) & RMS per beam\\
\hline
APEX-SZ & TBD & 1.0$\arcmin$ & 10 $\mu$K \\
SPT & 4000 deg$^2$ & 1.0$\arcmin$ & 10 $\mu$K \\
ACT & 100 deg$^2$ & 1.7$\arcmin$ & 2 $\mu$K \\
Planck & All-Sky & 7.1$\arcmin$ & 6.0 $\mu$K \\
\hline
\end{tabular}
\end{center}
\label{sz_inst}
\end{table}

There are also several centimeter
wave ($\approx$30 GHz) interferometers that are performing surveys of the SZE, such as the Arcminute Microkelvin Imager (AMI) \citep{ami} and
the Sunyaev-Zel'dovich Array (SZA) \citep{sza}. SZE surveys have the potential to
strongly constrain the $w$ parameter for dark energy, since they
sample clusters to large redshift.

Determining the abundance and distribution in mass and redshift of
massive galaxy clusters from observables is a challenging
exercise.  In the realm of cluster
abundance counts, one needs to know with high precision the mass range
of clusters probed in a flux-limited survey as a function of cluster
redshift. This determination depends on detailed knowledge of the
scaling between mass and light in clusters. It is critical to understand how cluster
observables correlate with cluster total mass in order to use clusters
of galaxies as precision cosmological tools.

\subsection{The Role of Simulations in Understanding Cluster Surveys}
Recent results indicate that high resolution N-body simulations
\citep{warren,heitman,reed} generate mass functions
which differ significantly from the Press-Schechter result and also from
the subsequent modifications of \citet{sheth} and
\citet{jenkins}. Since there is strong evidence that purely analytic
methods are insufficient, ``precision cosmology'' requires the use of
numerical simulations. In other words, in order to make predictions
which match the observed cluster population to percent-level
precision, analytic methods are inadequate. 

The output of numerical simulations of clusters can be compared to the
observed cluster mass function.  This comparison is non-trivial, however, due to
the uncertain nature of the conversion between observable quantities
(e.g., X-ray luminosity, SZE Compton y parameter, lensing shear) and cluster total
mass. Observations of the cluster gas typically depend on the detailed properties of
the hot baryons in clusters. It has been shown by our group and
others that cluster observables have a strong dependence on the baryonic physics in the ICM
\citep{hall06,nagai}, and are subject to an array of
uncertainties. However, the SZE signal integrated over the projected
cluster area (as defined in
Equation 2) inside $r_{500}$ is unique in that
the general scaling with mass is relatively independent of the assumed 
gas physics \citep{motl05,nagai}. While the normalization of the Y-M relation has some
dependence on ICM physics, the slope and tightness of the correlation
are unaffected \citep{nagai}. Most recent simulated light cone
calculations use the dark matter mass function generated by large N-body simulations like
the Hubble Volume simulation or the Millennium run, with
``painted on'' baryons to generate mock surveys \citep[e.g.,][]{evrard02,paint2}. These studies
typically assume the gas is isothermal and in hydrostatic equilibrium
with the dark matter potential. 

Both simulations including relevant physics \citep[e.g.,][]{white03, rasia06, hall06} and high resolution X-ray
observations of galaxy clusters \citep[e.g.,][]{utp_obs, 1E06, a168}
suggest that many clusters depart strongly
from both equilibrium and isothermality. These deviations
can have a strong impact on both the observable and derived properties
of clusters. Thus, in order to properly simulate sky surveys, it is
critically important to self-consistently include baryons in numerical
simulations. While some work has been done in this area
\citep[e.g.,][]{springel, white02, roncar, ronc07}, the largest volumes simulated were
small ($\sim$100-200 h$^{-1}$Mpc), and only sufficient to
generate synthetic light cones of roughly 1-4 deg$^2$. The simulation performed for this study
models a significantly larger physical volume than previous efforts,
has a higher peak physical resolution, and fully incorporates baryons
self-consistently. This allows us to perform much larger synthetic
surveys ($\sim$100 deg$^{2}$) than could be done with earlier
N-body/hydro simulations. 

Cosmological N-body/hydro simulations have advanced significantly in the last decade, such that the
simulation output now compares quite well to observations of galaxy
clusters \citep{sfw}. In particular, our group and others have shown that there is good agreement in simulated and
observed scaling relations between bulk
cluster ICM properties (e.g., cluster mass, X-ray luminosity, X-ray spectral
temperature) \citep{motlchar}. There remain important differences, particularly
for lower mass clusters, which indicates the need for a better
understanding of the details of baryonic cluster physics. The advance
of realism in simulations is a result of diligent, iterative efforts by various
groups of investigators to directly compare simulations to
observations. This study uses a large volume high resolution adiabatic
simulation, and serves as a template for more complex runs involving
additional non-gravitational physical processes which are likely important to
accurately modeling cluster surveys. These results should be
relatively robust in any case, as it has been shown that SZE survey
yields are relatively independent of cluster physics details \citep{white02}.

\subsection{Modeling SZE Surveys}
A variety of approaches have been taken to model SZE surveys. Most
involve either semi-analytic prescriptions or N-body
simulations where the gas is added in a post processing step
\citep[e.g.,][]{schulz}. As discussed in the previous section, there are limitations to
these methods, particularly in the assumptions of hydrostatic
equilibrium and isothermality. Some studies \citep[e.g.,][]{white02,holder}
have discussed the contribution of unresolved clusters/groups to the
signature of detected clusters. The presence of gas outside the cluster virial radii in low density
unbound structures such as filaments is potentially also quite important. This gas is completely absent in
non-hydrodynamic treatments of the simulation volume, appears
naturally in our calculation, and is expected to contain 40-50\% of
the baryons in the universe \citep{whim}. Since the SZE does not diminish with
distance, and results from a line of sight integral of the gas
pressure, the sum of all the flux from unbound gas could be a
significant contributor to the total flux in any cluster detection.

In this paper, we have examined the contribution of line-of-sight
baryonic gas to the expected SZE signal of simulated clusters. We
have stacked a (512 h$^{-1}$Mpc)$^3$ volume (comoving) adaptive mesh refinement (AMR)
N-body/hydro simulation to generate a survey of a light cone
subtending 100 square degrees on the sky. Our simulated
survey covers a larger sky area, and at higher angular resolution,
than any previous N-body + hydro simulated survey.  

There is extensive work in the literature on cluster detection
algorithms for SZE surveys,
\citep[e.g.,][]{diego,herranz,hobson,schafer,melin}. These methods
involve various techniques designed to spatially filter out the
primary CMB anisotropies (so called matched filtering), wavelet techniques, and application of public
tools such as SExtractor\footnote{http://terapix.iap.fr/rubrique.php?id\_rubrique=91/} \citep{sex}. We find the existing work to be quite
detailed, and do not introduce new algorithms of this type
here. Indeed, it is important to step back from the analyses which
have attempted to include all the relevant contaminants and
instrumental effects and explore the intrinsic difficulties resulting
from the cluster population as projected on the sky. There is a
limiting precision one can expect from cluster surveys irrespective of
the ability to remove instrumental effects, point source
confusion and sky backgrounds. This limit results from the possibly irreducible
confusion due to clusters, groups, lower mass halos and unbound gas, all
of which contribute to the SZE signal with a nearly identical spectral
signature. This study examines these effects with a more realistic
cosmological calculation than has typically been done, including the
full complement of baryons expected in the real universe. 

We explore the intrinsic limitations of SZE surveys by comparing and
characterizing the contribution of unresolved halos and unbound
filamentary gas to the cluster signal in samples that might result
from upcoming surveys
using a full hydro/N-body simulation of the cluster sky. In future
work, we will model backgrounds and instrumental characteristics as
has been done in the literature recently \citep[e.g.,][]{sehgal,
  melin, schafer} with a focus on techniques for accurately extracting
cluster properties and abundance. 

We discuss our methodology of simulating these large surveys in
Section 2, analytic predictions of SZE observables in Section 3,
present results in Section 4, discussion in Section 5, and summarize
our work in Section 6. 
\section{Methodology}\label{sec:Methodology}

\subsection{The Enzo Code}\label{sec:enzocode}

`Enzo'\footnote{http://lca.ucsd.edu/portal/software/enzo/} is a publicly available, extensively tested 
adaptive mesh refinement
cosmology code developed by Greg Bryan and colleagues \citep{bryan97,bryan99,norman99,oshea04,
2005ApJS..160....1O}.
The specifics of the Enzo code are described in detail in these papers (and references therein),
but we present a brief description here for clarity.

The Enzo code couples an N-body particle-mesh (PM) solver \citep{Efstathiou85, Hockney88} 
used to follow the evolution of a collisionless dark
matter component with an Eulerian AMR method for ideal gas dynamics by \citet{Berger89}, 
which allows high dynamic range in gravitational physics and hydrodynamics in an 
expanding universe.  This AMR method (referred to as \textit{structured} AMR) utilizes
an adaptive hierarchy of grid patches at varying levels of resolution.  Each
rectangular grid patch (referred to as a ``grid'') covers some region of space in its
\textit{parent grid} which requires higher resolution, and can itself become the 
parent grid to an even more highly resolved \textit{child grid}.  Enzo's implementation
of structured AMR places no fundamental restrictions on the number of grids at a 
given level of refinement, or on the number of levels of refinement.  However, owing 
to limited computational resources it is practical to institute a maximum level of 
refinement, $\ell_{max}$.  Additionally, the Enzo AMR implementation allows arbitrary 
integer ratios of parent
and child grid resolution, though in general for cosmological simulations (including the 
work described in this paper) a refinement ratio of 2 is used.

Since the addition of more highly refined grids is adaptive, the conditions for refinement 
must be specified.  In Enzo, the criteria for refinement can be set by the user to be
a combination of any or all of the following:  baryon or dark matter overdensity
threshold, minimum resolution of the local Jeans length, local density gradients,
local pressure gradients, local energy gradients, shocks, and cooling time.
A cell reaching
any or all of the user-specified criteria will then be flagged for refinement.  Once all 
cells of a given level have been flagged, rectangular solid boundaries are determined which 
minimally 
encompass them.  A refined grid patch is then introduced within each such bounding 
volume, and the results are interpolated to a higher level of resolution.

In Enzo, resolution of the equations being solved is adaptive in time as well as in
space.  The timestep in Enzo is satisfied on a level-by-level basis by finding the
largest timestep such that the Courant condition (and an analogous condition for 
the dark matter particles) is satisfied by every cell on that level.  All cells
on a given level are advanced using the same timestep.  Once a level $L$ has been
advanced in time $\Delta t_L$, all grids at level $L+1$ are 
advanced, using the same criteria for timestep calculations described above, until they
reach the same physical time as the grids at level $L$.  At this point grids at level
$L+1$ exchange baryon flux information with their parent grids, providing a more 
accurate solution on level $L$.  Cells at level $L+1$ are then examined to see 
if they should be refined or de-refined, and the entire grid hierarchy is rebuilt 
at that level (including all more highly refined levels).  The timestepping and 
hierarchy rebuilding process is repeated recursively on every level to the 
maximum existing grid level in the simulation.

Two different hydrodynamic methods are implemented in Enzo: the piecewise parabolic
method (PPM) \citep{Woodward84}, which was extended to cosmology by 
\citet{Bryan95}, and the hydrodynamic method used in the ZEUS magnetohydrodynamics code
\citep{stone92a,stone92b}.  We direct the interested reader to the papers describing 
both of these methods for more information, and note that PPM is the preferred choice
of hydro method since it is higher-order-accurate and is based on a technique that 
does not require artificial viscosity, which smoothes shocks and can smear out 
features in the hydrodynamic flow.

\subsection{Simulation Setup and Analysis}\label{sec:simsetup}

The simulation discussed in this paper is set up as follows.  We initialize
our calculation at $z=99$ assuming a cosmological model with  $\Omega_m = 0.3$, 
$\Omega_b = 0.04$, $\Omega_{CDM} = 0.26$, $\Omega_\Lambda = 0.7$, $h=0.7$ (in units of 100 km/s/Mpc), 
$\sigma_8 = 0.9$, and using an \citet{eishu99} power spectrum
with a spectral index of $n = 1$.  The simulation is of a volume of the 
universe 512~h$^{-1}$~Mpc (comoving) on a side with a $512^3$ root grid.  The dark matter
particle mass is $7.228 \times 10^{10}$~h$^{-1}$~M$_\odot$ and the mean baryon mass resolution
is $1.112 \times 10^{10}$~h$^{-1}$~M$_\odot$.  The simulation was then evolved to $z=0$ with
a maximum of $7$ levels of adaptive mesh refinement (a maximum spatial resolution of 
$7.8$~h$^{-1}$ comoving kpc), refining on dark matter 
and baryon overdensities of $8.0$ (to ensure an approximately Lagrangian mass
resolution in baryonic structures).  The equations of hydrodynamics were
solved with the Piecewise Parabolic Method (PPM) using the dual energy formalism.
The entire grid hierarchy (including both particle and baryon information) was 
written out at regular intervals, and in particular, data was output at intervals
of $\Delta z = 0.25$ between $z=3$ and $z=2.5$ (inclusive), and $\Delta z = 0.1$ between $z=2.5$ 
and $z=0.1$ (inclusive).

Analysis was performed on every data output between $z=3$ and $z=0.1$ in an
identical way.  The HOP halo-finding algorithm \citep{eishut98} was applied to the 
dark matter particle distribution to produce a dark matter halo catalog.
Spherically-averaged, mass-weighted radial profiles of various baryonic and dark matter
quantities including density, temperature, and pressure were then generated 
for every halo in the catalog with an
estimated halo mass greater than $4 \times 10^{13}$~M$_\odot$.  These radial 
profiles were used to calculate more accurate virial masses and radii as well 
as an estimate for the Compton $y$ parameter as a function of impact parameter on
the halo.  Projections of the integrated Compton $y$ value along the line of sight were created
for each of the three axes along the simulation volume, with two projections per axis --
one of the front half of the simulation volume, and one along the back half.  Each
projection has an approximate depth of $\Delta z = 0.1$.  These projections have a 
resolution of $2048$ pixels on a side.

\subsection{Generation of the ``Santa Fe'' Light Cone}\label{sec:lightcone}

Mock SZE observations of a $10^{\circ} \times 10^{\circ}$ patch of sky are 
generated by stacking projections from the simulation discussed in 
Section~\ref{sec:simsetup}.  These ``light cones'' are created by stacking 
projections of the Compton $y$ parameter at each redshift output.  At each redshift, the
 projection is chosen to be along a random axis, and has been randomly shifted in space
such that the positions of large scale structure is uncorrelated.  Additionally, 
the projections have been rescaled to the resolution of the light cone, which is 2048 pixels
per side, or a resolution of $17.58$ arc seconds per pixel.  This scaling may involve 
tiling (for redshifts
where 512 h$^{-1}$ Mpc comoving corresponds to less than $10^{\circ}$ on the sky) or
interpolating (for redshifts where  512 h$^{-1}$ Mpc comoving corresponds to more
 than $10^{\circ}$ on the sky). Secondary maps are created which
 include only the
Compton $y$ parameter contributed by gas within the virial radius of halos with
masses above $5 \times 10^{13}$~M$_\odot$, and only gas outside of the virial radius
of these objects.  200 of these mock ``light cones'' at this size and
resolution are created using different random 
seeds.  These light cones (named ``Santa Fe'' light cones due to the location
where the project was conceived) have angular resolution which
is significantly higher than any current or proposed SZE observational
campaign. \textit{The goal of this
analysis is not to determine an optimal method for source finding, but
to determine the contamination from unresolved halos and unbound gas for a simple method.} 

\section{Analytic Predictions for SZE Observables}\label{sec:theory}
Here we describe some of the theory behind the use of SZE cluster
observations in constraining cosmology. Though we are aware that these
types of analytic calculations have been performed previously, we show
them here to motivate not just the current analysis, but that which
will be performed for subsequent papers in this series. 
 
One of the most useful methods for retrieving cosmological information from
SZE observations of galaxy clusters is by the calculation of galaxy cluster
counts as a function of redshift.  The number of galaxy clusters above some
given minimum mass $M_{min}(z)$ in a redshift bin of width $dz$ and solid angle 
$d\Omega$ can be defined using the Press-Schechter formalism 
\citep{ps} as
\begin{equation}
\frac{dN}{dz d\Omega}(z) = \frac{dV}{dz d\Omega}(z) \int_{M_{min}(z)}^{\infty} dM \frac{dn}{dM}(M,z)
\label{eqn-Nofz}
\end{equation}
where $dV/dzd\Omega$ is the cosmological comoving volume element at a given redshift, 
and $\frac{dn}{dM} dM$ is the comoving halo number density as 
a function of mass and redshift.  The latter is expressed as by \citet{jenkins} as
\begin{eqnarray}
\frac{dn}{dm} (M,z) & = & -0.315 \frac{\rho_0}{M} \frac{1}{\sigma_M} \frac{d\sigma_M}{dm} \nonumber\\
& exp & \left[ -|0.61-log(D(z)\sigma_M)|^{3.8} \right]
\label{eqn-dndm}
\end{eqnarray}
where $\sigma_M$ is the RMS density fluctuation, computed on mass scale M from the $z=0$ linear power spectrum \citep{eishu99}, 
$\rho_0$ is the mean matter density of the universe, defined as $\rho_0 \equiv \Omega_m \rho_c$ (with $\rho_c$ being 
the cosmological critical density, defined as $\rho_c \equiv 3 H_0^2 / 8 \pi G$), and $D(z)$ is the linear growth function, given by this fitting function:
\begin{eqnarray}
& &D(z) = \frac{1}{1+z} \frac{5 \Omega_m(z)}{2} \nonumber\\
& &\left\{ \Omega_m(z)^{4/7} - \Omega_\Lambda(z) + [1+\frac{\Omega_m(z)}{2}][1+\frac{\Omega_\Lambda(z)}{70}]\right\}^{-1}
\label{eqn-DofZ}
\end{eqnarray}
\citep{1992ARA&A..30..499C}, with $\Omega_m(z)$ and $\Omega_\Lambda(z)$ defined as:
\begin{equation}
\Omega_m(z) = \Omega_{m,0} (1+z)^3 E^{-2}(z)
\label{eqn-omofz}
\end{equation}
and
\begin{equation}
\Omega_\Lambda(z) = \Omega_{\Lambda,0} E^{-2}(z)
\label{eqn-olofz}
\end{equation}
where $\Omega_{m,0}$ and $\Omega_{\Lambda,0}$ are the density of matter and cosmological constant at the
present day, expressed in units of the critical density.  The cosmological volume element is given by:
\begin{equation}
\frac{dV}{dz d\Omega}(z) = \frac{c}{H_0} \frac{ (1+z)^2 D_A^2}{E(z)}
\label{eqn-dV}
\end{equation}
where $D_A(z)$ is the angular diameter distance as a function of redshift, c is the speed of light, 
$H_0$ is the Hubble parameter at $z=0$, and E(z) is given by:
\begin{equation}
E^2(z) = \Omega_{m,0} (1+z)^3 + \Omega_\Lambda
\label{eqn-Eofz}
\end{equation}
in a flat universe with a cosmological constant \citep{peebles-book}.  The RMS amplitude of the 
density fluctuations as function of mass
at any redshift, as smoothed by a spherically symmetric window function with a characteristic comoving
radius R, can be computed from the matter power spectrum using the relation:
\begin{equation}
\sigma^2(M,z) = \int_0^\infty \frac{dk}{k} \frac{k^3}{2 \pi^2} P(k,z) |\mbox{\~{W}}_R(k)|^2
\label{eqn-sigma}
\end{equation}
where $\mbox{\~{W}}_R(k)$ is the Fourier transform of the real-space top hat smoothing function:
\begin{equation}
\mbox{\~{W}}_R(k) = \frac{3}{k^3 R^3} \left[ sin(kR)-kRcos(kR)\right]
\label{eqn-W}
\end{equation}

The radius R is calculated for a given mass by using the relation $M = \frac{4}{3} \pi R^3 \Omega_m \rho_c$,
and $\sigma(M,z)$ is normalized to $\sigma_8$, defined as the RMS density fluctuation when smoothed by a sphere with a comoving radius of $8$~h$^{-1}$ Mpc at $z=0$, using observations
of large-scale structure or the cosmic microwave background. The matter power spectrum is expressed as:
\begin{equation}
\frac{k^3}{2 \pi^2} P(k,z) = \left( \frac{c k}{H_0} \right)^{3+n} T^2(k) \frac{D^2(z)}{D^2(0)}
\label{eqn-pofk}
\end{equation}
where $T(k)$ is the matter transfer function which describes the way in which the processing of the 
initial spectrum of matter density fluctuations during the radiation-dominated era~\citep{peebles-book} and
$D(z)$ is the fitting function for the linear growth function, as given in Equation~\ref{eqn-DofZ}.
In the calculations discussed in this paper, we use the transfer function $T(k)$ provided by \citet{eishu99}. 

\begin{figure}
\begin{center}
\includegraphics[width=0.45\textwidth]{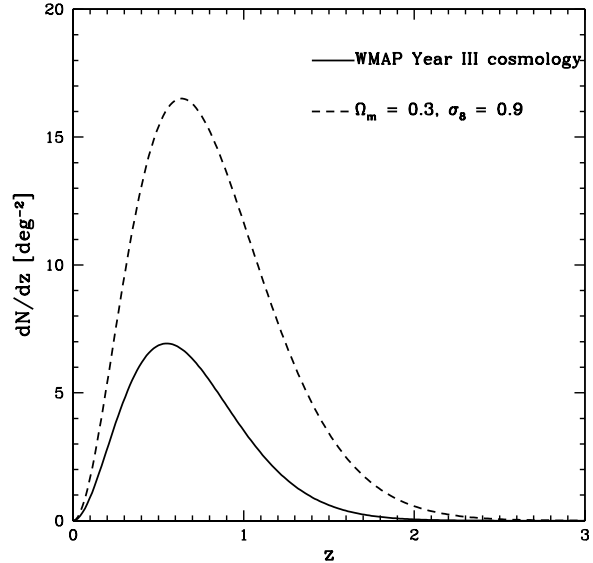}
\end{center}
\caption{Number of galaxy clusters per square degree as a function of redshift with 
$M_{tot} \geq 1 \times 10^{14}$~h$^{-1}$~M$_\odot$ in the WMAP Year III ``most favored''
cosmology (solid line; $\Omega_m = 0.268$,~$\sigma_8 = 0.776$) and the cosmology used in the simulation in this paper
(dashed line; $\Omega_m = 0.3$,~$\sigma_8 = 0.9$).  The distribution peaks at 
$z \simeq 0.55$ for the WMAP Year III cosmology and at $z \simeq 0.64$ for the cosmology used in the simulations in this paper.
}
\label{fig.dNdz}
\end{figure}

Figure~\ref{fig.dNdz} shows the number of galaxy clusters per square degree as a function of redshift with
$M_{tot} \geq 1 \times 10^{14}$~h$^{-1}$~M$_\odot$ in the WMAP Year III ``most favored''
cosmology ($\Omega_m = 0.268$,~$\sigma_8 = 0.776$) and in the cosmology used in the simulation in this paper
($\Omega_m = 0.3$,~$\sigma_8 = 0.9$).  Due to the higher $\Omega_m$ and $\sigma_8$, significantly more 
galaxy clusters are expected to be seen in this simulation than one would expect given the WMAP Year III result.

\begin{figure}
\begin{center}
\includegraphics[width=0.45\textwidth]{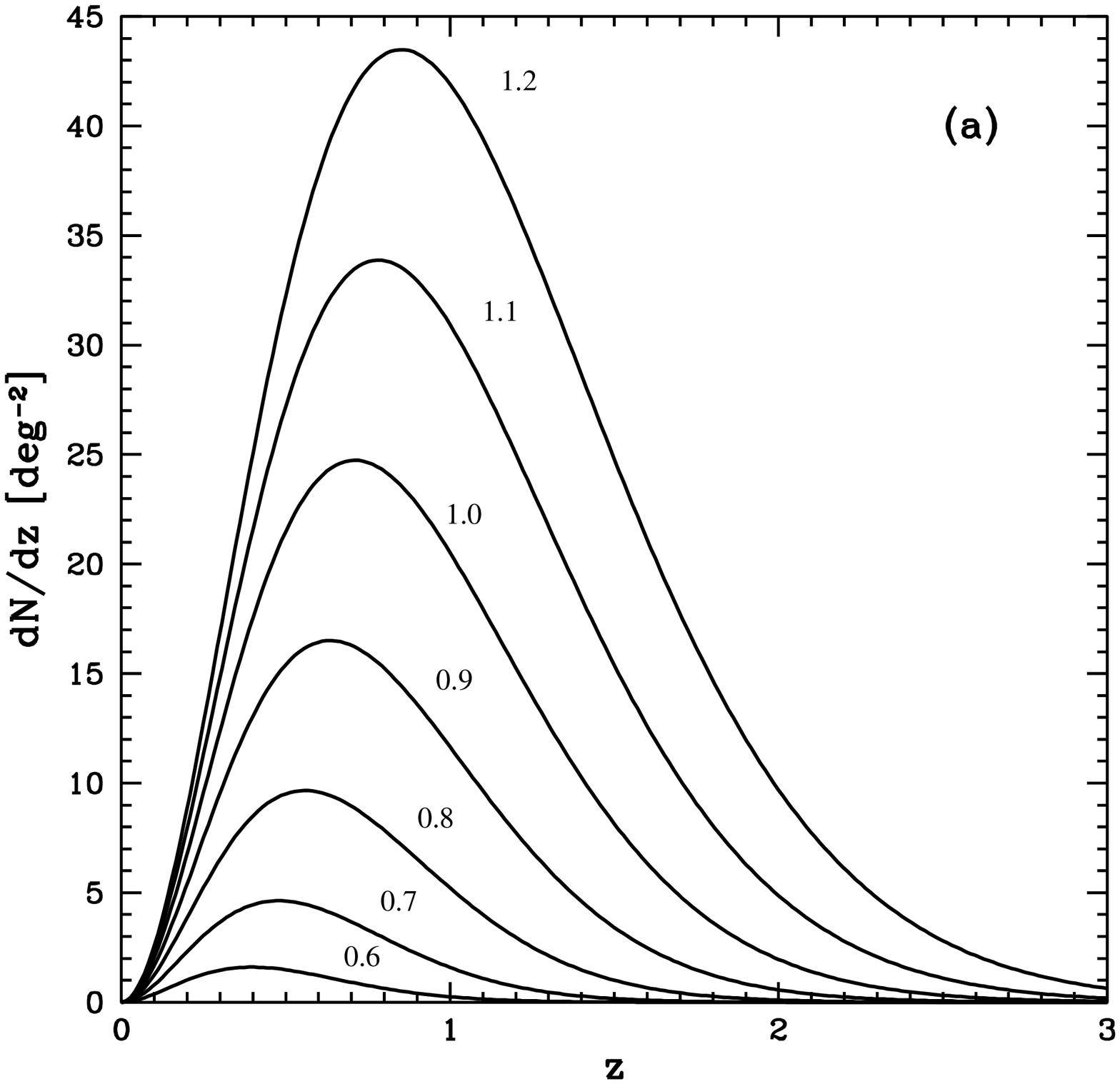}
\includegraphics[width=0.45\textwidth]{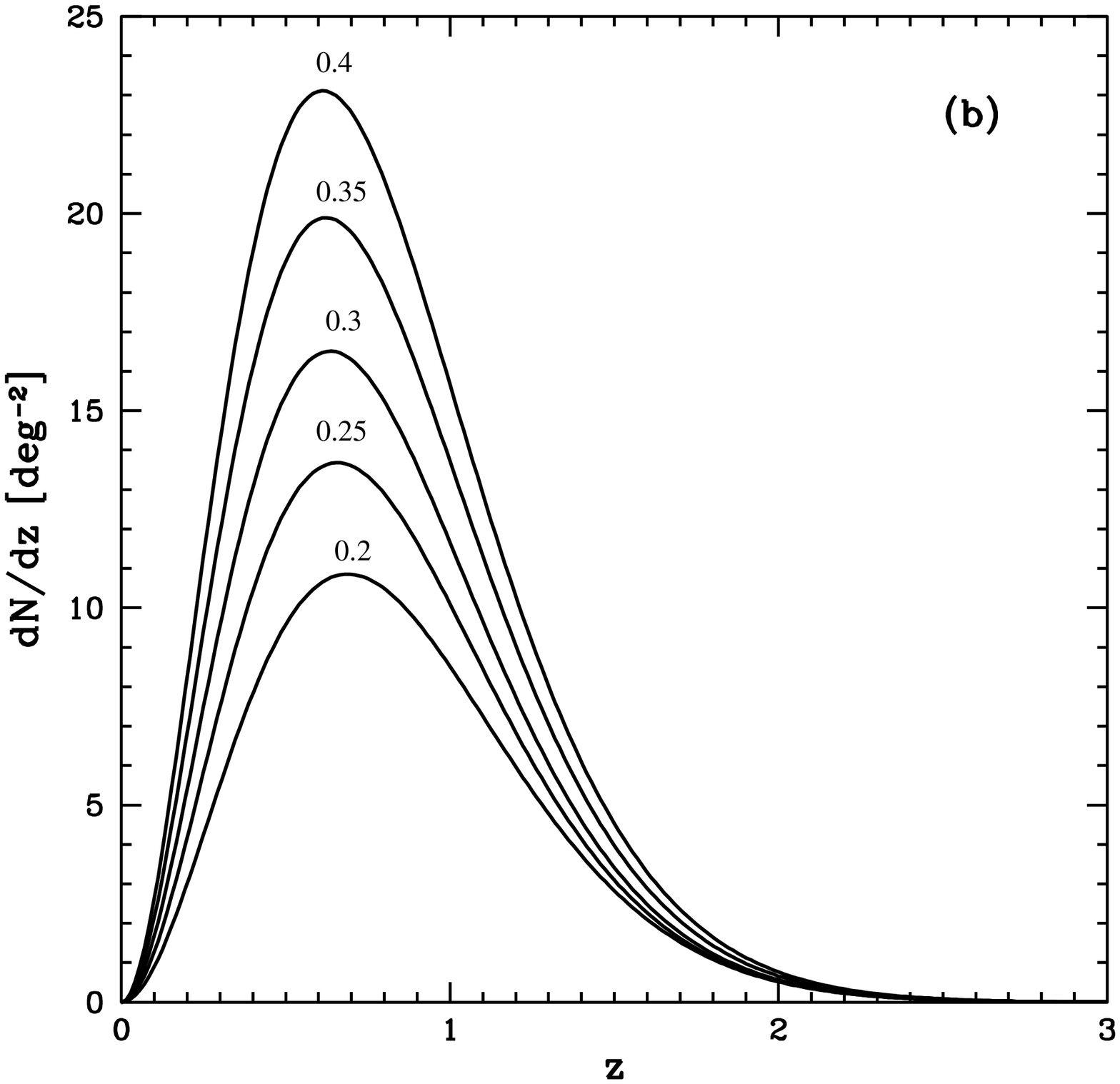}
\end{center}
\caption{Number of galaxy clusters per square degree as a function of redshift with 
$M_{tot} \geq 1 \times 10^{14}$~h$^{-1}$~M$_\odot$ for a variety of cosmologies.
Panel (a):  All  parameters except $\sigma_8$ are held constant, using the same
cosmological parameters as in the simulation described in this paper ($\Omega_m = 0.3$, $\Omega_b = 0.04$, 
$\Omega_\Lambda = 0.7$, $h=0.7$, $n=1$), but varying $\sigma_8$ from $0.6$ to $1.2$ in steps of $0.1$ (bottom to top lines).
Panel (b):  All parameters except $\Omega_m$ and $\Omega_\Lambda$ are held constant using the same cosmological parameters
as in the simulation described in this paper ($\Omega_b = 0.04$, $h=0.7$, $n=1$, $\sigma_8 = 0.9$), but varying $\Omega_m$
from 0.2 to 0.4 in steps of 0.05 (bottom to top lines) and keeping $\Omega_\Lambda = 1 - \Omega_m$.
}
\label{fig.dNdz-cosmo}
\end{figure}

Figure~\ref{fig.dNdz-cosmo} shows the number of galaxy clusters per square degree as a function of redshift
with $M_{tot} \geq 1 \times 10^{14}$~h$^{-1}$~M$_\odot$ for a variety of cosmological models.
Panel (a) shows a sequence of cosmologies where all  parameters except $\sigma_8$ are held constant, using the same
cosmological parameters as in the simulation described in this paper ($\Omega_m = 0.3$, $\Omega_b = 0.04$, 
$\Omega_\Lambda = 0.7$, $h=0.7$, $n=1$), but varying $\sigma_8$ from $0.6$ to $1.2$ in steps of $0.1$ (bottom to top lines).
Panel (b) shows a sequence of cosmologies where all parameters except $\Omega_m$ and $\Omega_\Lambda$ are held constant 
using the same cosmological parameters
as in the simulation described in this paper ($\Omega_b = 0.04$, $h=0.7$, $n=1$, $\sigma_8 = 0.9$), but varying $\Omega_m$
from 0.2 to 0.4 in steps of 0.05 (bottom to top lines), and keeping $\Omega_\Lambda = 1 - \Omega_m$.  It is interesting
to note that varying $\sigma_8$ while holding all other cosmological parmaters results in a change in both the overall
number of halos (a
factor of more than 3 between WMAP and the cosmology in our
simulation) and the redshift at which the distribution peaks .
Both effects can be explained by examining the term in the 
exponent in equation~\ref{eqn-dndm}.  The comoving number density of halos is maximized when $0.61-log(D(z)\sigma_M) = 0$,
or $D(z) \times \sigma_M \simeq 4.07$.  For a given set of cosmological parameters, increasing $\sigma_8$ increases $\sigma_M$ for 
a given mass value, and thus maximizes the number density of halos at a given mass at a smaller linear growth factor (or, more
intuitively, a higher redshift).  Variation in $\Omega_m$ and $\Omega_\Lambda$ while holding all other parameters constant
effectively results in a change in the normalization of the overall halo number density, while keeping the redshift at which
this distribution peaks roughly constant.

These results are of great interest for upcoming SZE surveys, which
will sample clusters to relatively low mass at high redshifts (z =
0.5-1.0) compared to optical and X-ray surveys. This redshift range is
where we should expect the largest difference between a low $\sigma_8$
cosmology preferred by WMAP and a higher value typically used in
cosmological simulations. The large difference between the abundance
of clusters in these different cosmologies should lead to large
differences in the number of identified clusters in surveys. This
means that very early on in any SZE survey it should become fairly
obvious which cosmology is preferred by cluster observations. This
will be an even stronger constraint when optical follow up
observations are used to
determine redshifts, which will break the degeneracy between $\Omega_M$ and
$\sigma_8$.

\section{Results}\label{sec:results}

\subsection{Simulated Mass Function}\label{sec:massfctn}

\begin{figure}
\begin{center}
\includegraphics[width=0.45\textwidth]{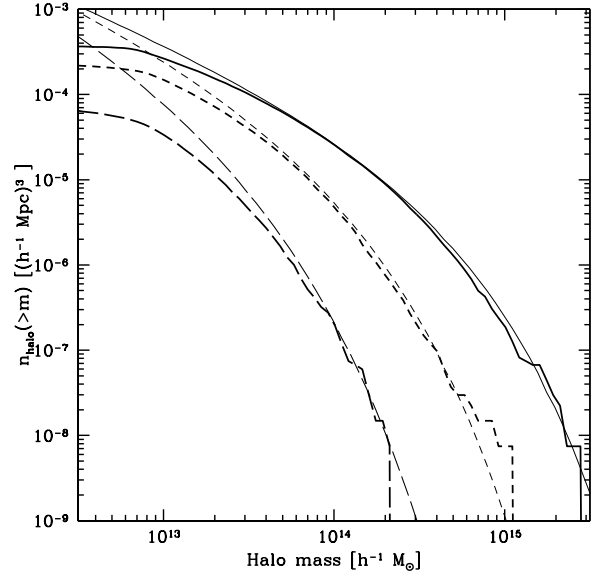}
\end{center}
\caption{Cumulative number density at several redshifts.  Solid: $z=0.1$.  Short-dashed: $z=1.0$.  Long-dashed: $z=2$.
 Thick lines are simulation results calculated using the halo virial masses (dm+gas) and the thin
lines are the Warren fitting function \citep{warren}.
}
\label{fig.cumnumdens}
\end{figure}

One of the most basic tests of the correctness of a cosmological simulation 
is whether or not it can match the predicted halo mass function for a given 
cosmology.  This is particularly important in the context of creating simulated
sky maps for cosmological surveys of any kind, given that the number of 
halos and their redshift distribution is the most basic test of cosmological 
properties.  Figure~\ref{fig.cumnumdens} shows the cumulative number density
of cosmological halos as a function of mass for several redshifts which span the
range of interest for the topics discussed in this work ($z = 0.1 -3$).  
The halos were found as described in Section~\ref{sec:simsetup}, and 
the masses used are the total
halo virial mass (including both baryons and dark matter), rather than the mass returned
by the Hop halo finding algorithm~\citep{eishut98}.  We also show the fitting function 
for cumulative halo number density obtained by \citet{warren} as a reference.  At 
low redshifts ($z=0.1-1$) the mass function of halos from the simulation agrees quite well with 
the fitting function over the mass range of interest.  This is encouraging, as the bulk of 
galaxy clusters in the universe are at
these low redshifts (as shown in Figure~\ref{fig.dNdz}).  At higher redshifts ($z=2$), the 
fitting function and halo mass function only agree at the highest masses 
($M_{halo} \geq 10^{14}h^{-1}$~M$_\odot$).  This is to be expected: grid-based codes, including adaptive
mesh codes, tend to suppress low-mass halo formation, particularly at high redshift, as has
been seen in recent code comparisons~\citep{2005ApJS..160....1O,
  2005ApJS..160...28H}. This suppression in our study only exists due
to the choice of simulation setup. In order to model such a large physical
volume with both N-body + hydrodynamics, we must sacrifice mass
resolution due to computational effort concerns. Given
that the cosmological surveys of interest will only be sensitive to halos in the mass range
where this simulation agrees well with the Warren et al. fit, and the relative paucity of these
halos at $z \ga 2$, there is little cause for concern.

\subsection{SZE Angular Power Spectrum}\label{sec:szpspec}
We have calculated angular power spectra for the 200 survey images to
determine the cosmic variance in this field, and the general form of
the power spectrum. The calculation involves determining the
power spectral density of the image as a function of image
scale. Additionally, we have performed the same calculation on maps
smoothed with the beam size at 144 GHz for each of four upcoming SZE
survey instruments. We have modified the images to model the limitations
of these instruments in a simple way.  First, for each instrument, we
have gaussian smoothed the image with the FWHM of the beam at a wavelength
(frequency) of 2.1mm (144 GHz) corresponding to the maximum SZE
decrement. All the upcoming SZE single-dish surveys will have the
capability of operating at this wavelength. We have also rebinned each
image such that the beam diameter is represented by at least two pixels in
order to be Nyquist sampled. A ``background'' in each case is
generated by adding a gaussian distributed variation with a FWHM equal
to each survey's stated limiting sensitivity. 

\begin{figure*}
\plotone{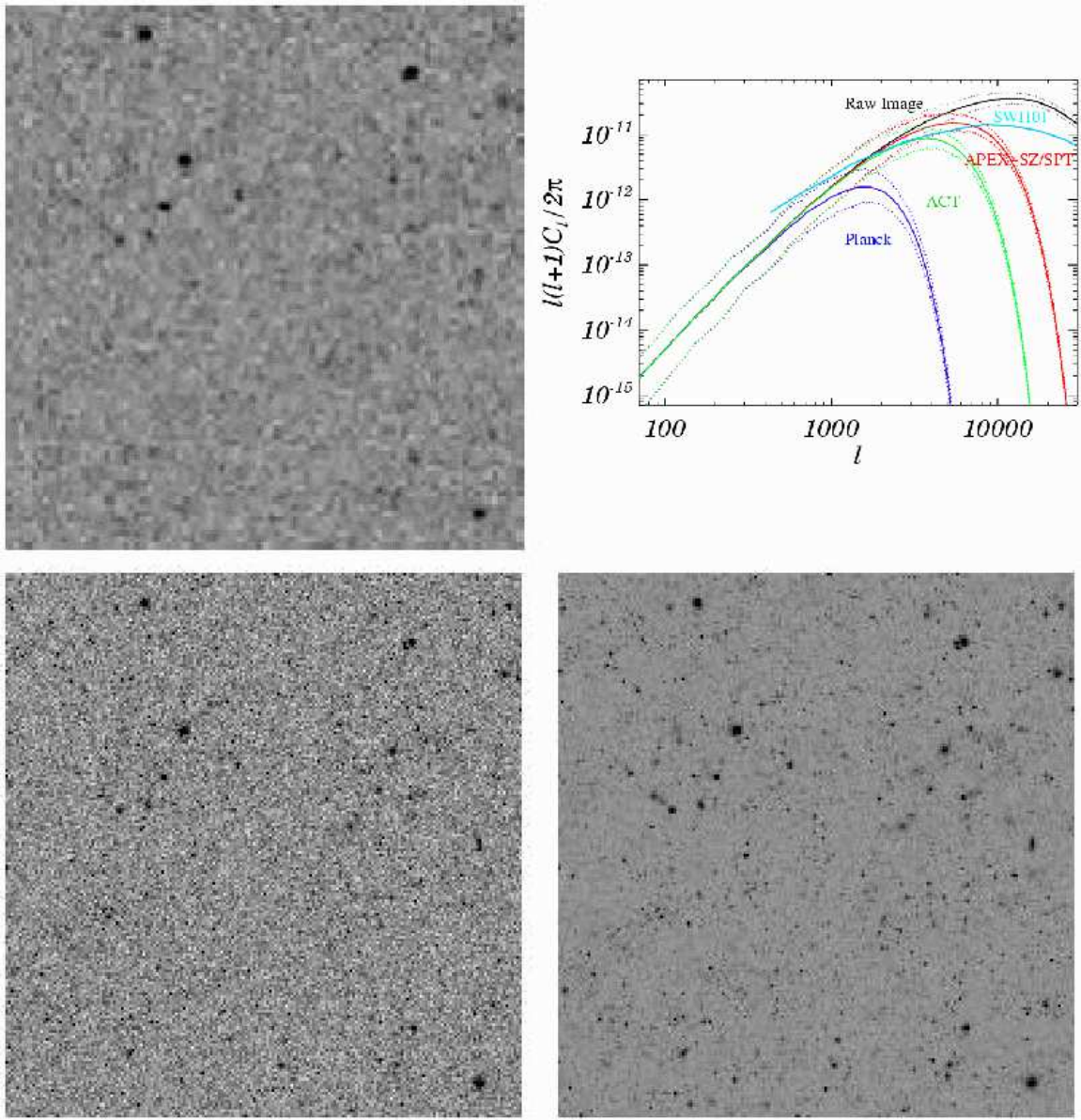}
\caption{Upper Left: Light cone survey image of 100 square degrees
  modified to simulate the beam size and limiting sensitivity of the
  Planck Surveyor all-sky survey at 144 GHz (7.1$\arcmin$,
  6$\mu$K). Upper right: Angular power spectra generated from these images. Lower left: Light cone image modified to
  simulate APEX-SZ/SPT survey characteristics at $\sim$144 GHz
  (1.0$\arcmin$, 10$\mu$K). Lower right: Light cone image modified to
  simulate ACT survey (1.7$\arcmin$, 2$\mu$K). Numbers in parentheses
  indicate (beam size, survey sensitivity/beam) at 144 GHz.}
\label{images}
\end{figure*}
The result of our analysis of the power spectrum of the simulated SZE
surveys is shown in Figure \ref{images}. The solid lines are the
mean values of the power from the 200 stacking realizations of the
light cone, and the dotted lines indicate the range in which 90\% of
our simulated light cone power spectra fall. The color indicates
which survey's characteristics were used to generate the result. 

The location of the peak of each curve is a function of the resolving
power of the survey, as is its amplitude, in that the power at the
smallest angular scale for each survey is different.  The cyan line
(labeled SWH01) is the result obtained by \citet{springel} with a 1 degree
angular scale light cone generated from an SPH N-body/hydro
simulation run with cosmological parameters matching our simulation. The raw power spectrum is generated with no smoothing from
the raw light cone survey images. The SWH01 result is different from
ours, possibly a result of lower spatial resolution in the simulation
(resulting in a deficit in the small scale angular power) and had a smaller
angular field (1 square degree) than ours, leading to a higher $\ell$
cutoff ($\ell$=400) than we show ($\ell$=70). As has been shown
previously, and specifically in Figure 4 of \citet{springel}, analytic
predictions of the SZE power vary widely depending on the calculation
method. They neither agree with the simulation results, nor in many
cases with each other. It seems that the analytic result has not
converged, and so to avoid confusion we do not plot it in this work.  

It is clear that each survey will sample a slightly different range of scales, though
they obviously all are able to measure the large angular scale power. Each power
spectrum peaks and turns over where the angular scale of the beam
begins to limit the measurement at high $\ell$. The power spectrum is not very sensitive to
non-gravitational gas physics at low multipole numbers, but primarily
is sensitive to cosmology, particularly the value of $\sigma_8$
\citep{white02,holder}. The addition of non-gravitational physics does impact
the small scale power however, for example \citet{holder} show that
preheating results in reduced small scale power in their simulated
images. 

What is also of interest in this analysis is the size of the variance shown
by the 90\% range error bars. On 100 square degree patches of the sky
at $\ell \la 2000$ the deduced power can be different by factors of
5-8. This indicates that the power spectrum can be quite different
from one area of the sky to another, and clearly requires greater sky
coverage to be well constrained. The cosmic variance range does
not become very small until $\ell \ga $ a few thousand. 

\subsection{SZE Source Identification}\label{sec:szsourceid}
\begin{figure}
\includegraphics[width=0.45\textwidth]{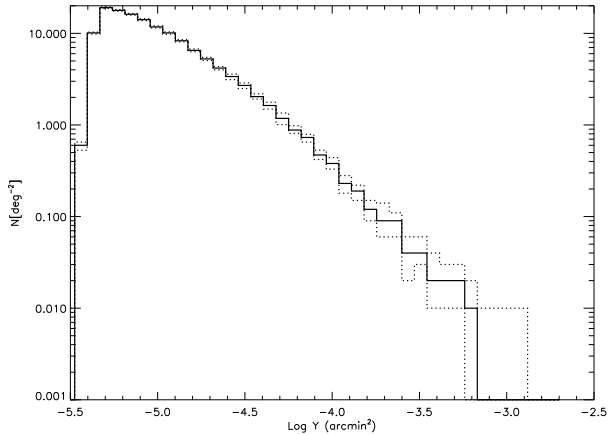}
\caption{Plot of number of objects per square degree in the simulated
  survey image as a
  function of integrated Compton $y$ parameter. Dotted lines are 90\% variance as calculated from
  200 stacking realizations of the light cone. }
\label{logNS}
\end{figure}

To identify objects in the light cone images, we simply locate the
projected clusters from the three-dimensional halo finding in the
image plane. Since we design the shifting and stacking strategy, it is
trivial to determine the image plane location of each cluster in each
redshift slice of the light cone. We also have calculated the
spherically averaged radial profiles projected into the image plane of
the Compton $y$-parameter. For each cluster with M $\geq 10^{14}
M_{\odot}$, we can then calculate the integrated value of Y (when
comparisons to analytic results are not performed, we use true masses
on the simulation grid, without the $h^{-1}$ modifier). The
result is shown in Figure \ref{logNS}. In this case, we have
integrated out to the virial radius of each cluster. We show the
variance in 200 stacking realizations of the simulated light cone,
indicating the increase in variation at high flux, as expected due to
rare, very massive objects projected into some fraction of the
images. 

\begin{figure}
\includegraphics[width=0.5\textwidth]{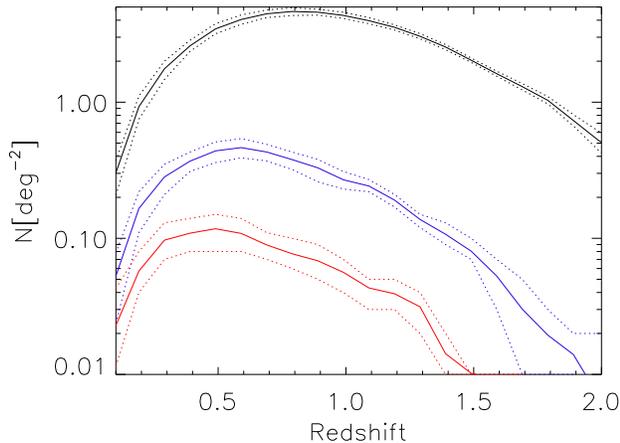}
\caption{Angular density of clusters in the light cone images as a
  function of redshift. Black solid line is the redshift distribution
  of clusters with M$\geq 10^{14} M_{\odot}$. Blue solid line is for
  M$\geq 3.0 \times 10^{14} M_{\odot}$, red is for M$\geq 5.0\times 10^{14} M_{\odot}$. Dotted lines are
  90\% variance of 200 independent stacking realizations of the light cone.}
\label{fzplot}
\end{figure}

We are also interested in the redshift distribution of the objects,
which can be compared to analytic estimates. 
The result of this analysis is shown
in Figure \ref{fzplot}. In each case, we have taken the mean value and
plotted it as a solid line, and the 90\% variance in 200 light cone
realizations as dotted lines. The black lines indicate all clusters
with total masses above $10^{14} M_{\odot}$ in the simulation which are in the projected
field of the survey. Blue and red lines are for clusters above higher
mass cutoffs, blue for M$\geq 3.0 \times 10^{14} M_{\odot}$ and red
for M$\geq 5.0\times 10^{14} M_{\odot}$. These give a rough indication of
the expected redshift distribution of identified clusters in upcoming
surveys.  

\begin{figure}
\includegraphics[width=0.5\textwidth]{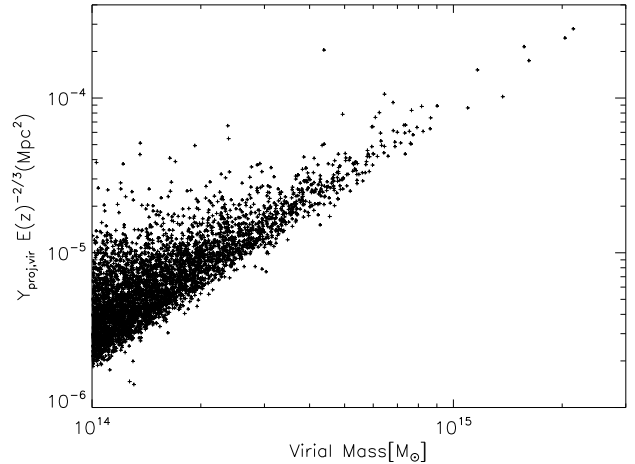}
\caption{Projected integrated Y value plotted against cluster mass. Y
  is integrated from the simulated light cone survey images from the
  center of each cluster out to the projected virial radius. 
  Y is converted from angular units using the angular diameter
  distance appropriate to the redshift of the matched simulated
  cluster. Y is also scaled with E(z)$^{-2/3}$ to account for the
  cosmological variation of the mass scaling relation.}
\label{yvm}
\end{figure}

\begin{figure}
\includegraphics[width=0.5\textwidth]{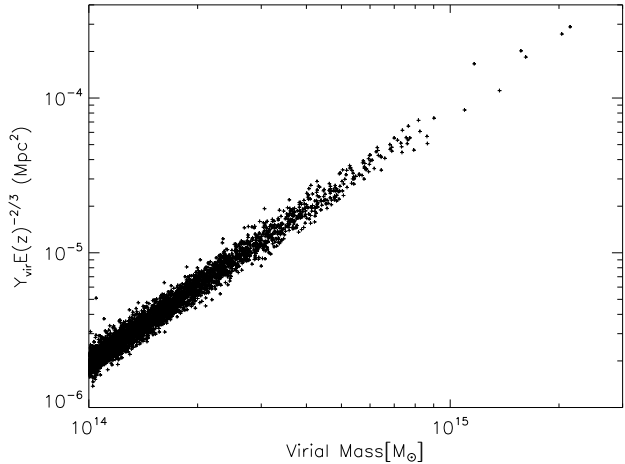}
\caption{Integrated Compton $y$ vs Total mass relationship extracted
  directly from the simulation data. Y is the projected integrated SZE
  $y$ parameter from each cluster out the the virial radius, total mass is
  from the simulation grid for each object. Y is also scaled with E(z)$^{-2/3}$ to account for the
  cosmological variation of the mass scaling relation.}
\label{yvmtrue}
\end{figure}

For one projected light cone image, we show the
Y vs M relationship in Figure \ref{yvm}. In this case, $Y$ is
calculated by integrating the value of Compton $y$ in the image out to
each cluster's \textit{projected} virial radius. The value of $Y$ is corrected for redshift since it
depends on E(z)$^{-2/3}$ \citep[see][]{nagai} due to the cosmological
dependence of the cluster M-T relation, and the angular scale is converted to Mpc
through use of the value of $D_A$ for each cluster. Contrast this plot with Figure \ref{yvmtrue}, for a single
realization of the light cone image, where the true projected Y is plotted against
cluster true mass from the simulation. The true value for Y is
calculated from the projected spherically averaged radial profiles of
each simulated cluster, but includes the gas out only to the
virial radius \textit{in three dimensions}. This true relation has
extraordinarily tight scatter, as has been shown previously
\citep{dasilva, motl05, nagai}. The difference in the two plots is
effectively the difference between the cluster's true integrated SZE
and the SZE integrated in a cylinder with radius equal to the cluster's
virial radius. For a narrow mass bin around 3.0$\times$10$^{14}
M_{\odot}$, the median bias in Y due to projection is 79\%, and the
scatter is +32\%/-16\% about that median, a significant increase over
the scatter in the true Y-M relationship. Note that some clusters in Figure \ref{yvm} appear to
have lower value of Y in projection than the true value for that
cluster. These are the clusters which lie near the edge of the
simulated survey image, and the cluster extends beyond the image
edge. As identified by other studies,
errors in extracting the correct value of Y should dominate the error
budget for this relation \citep[e.g.,][]{white02,melin}. These figures
illustrate the difficulty in accurately estimating Y from observations. A similar result is noted by
\citet{white02} using data from a smoothed particle hydrodynamics (SPH)
simulation. We have corrected the primary object in the
source region for cosmological evolution in the M-T relation, so the
remaining scatter results from line of sight projection effects. This
means that there are secondary bound objects (and unbound gas) which are projected
into the cluster's source region as will be described in the next
section. 
 
\subsection{Confusion Problems}
In this section we address the confusion resulting from clusters,
groups, and smaller mass halos, as opposed to confusion resulting from
radio and millimeter wave point sources in the cluster fields. This
work also includes the additional flux contributed by unbound gas, which contributes at some level in
the real universe, but has been ignored in most simulations of SZE
surveys. While there are several definitions of
confusion in the literature, even for SZE surveys, we have chosen to
define cluster confusion as the number of true cluster- or group-mass objects
in the simulation whose centers are within the source
region in projection. Since upcoming SZE surveys are unlikely to
detect cluster gas out to the virial radius, we have chosen a smaller
radius ($r_{500}$) as the source region. For each of the upcoming
surveys, we have defined the source region as a radius of a full beam
diameter, presuming that if two objects were imaged by two separate
non-overlapping beams, that there is a possibility they would not be
confused. This also presumes that the secondary object is bright
enough to be detected on its own. One can also define confusion as an error in recovered
flux \citep{holder} from sources found by progressive matched filtering at
different angular scales corresponding to variation of the angular size
of clusters as a function of redshift \citep[e.g.,][]{melin}. 

\begin{figure}
\includegraphics[width=0.5\textwidth]{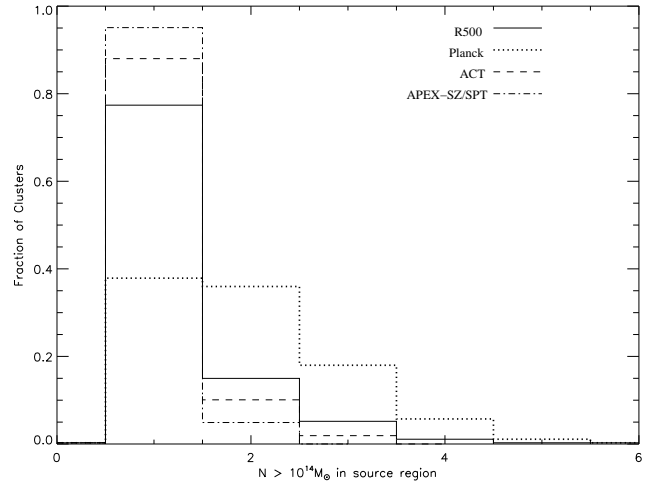}
\caption{Histogram of number of simulated clusters with M $>10^{14}
  M_{\odot}$ per source region for each of survey modified light
  cones. Source region is defined as inside $r_{500}$ projected for the solid
  line, and as within 1 beam diameter (at 144 GHz) distance for each of the survey
  instruments listed. Cluster identifications for all 200 light cones are tallied
  and used in the fraction.}
\label{all_source}
\end{figure}

Whether this type of confusion can be mitigated
depends on a variety of factors, including the mass of the additional
secondary objects in the source region, redshift distribution and
angular scale of the objects and the observing beam and multiple
wavelength identfication of the objects (e.g., optical, X-ray,
lensing). We endeavor here then simply to characterize the amount of
said confusion, leaving the mitigation of this problem to future
work. Figure \ref{all_source} shows for each of the surveys considered
a histogram of the number of objects above 10$^{14}
M_{\odot}$ in the source regions (as defined above) of all clusters
with M $\geq 3.0\times10^{14} M_{\odot}$. These mass limits are chosen
somewhat arbitrarily, though for the ground-based surveys,
$3.0\times10^{14} M_{\odot}$ is close
to the expected mass limit. $10^{14} M_{\odot}$ is chosen for the
contaminating objects on the expectation that one or more object above
that mass in the source region should lead to a significant bias in
the SZE flux from that expected from a single cluster in the
detectable mass range. As in Section 4.3, we have identified each
cluster by its projected image plane position associated with its true
three-dimensional location in the simulation after stacking.  It is clear that for Planck, there are a
high number of objects per source region above this relatively high
mass. In fact, roughly 40\% of source regions have only one object above
this mass cut, while nearly 60\% have more than one. Contrast this
result with that of APEX/SPT, where 95\% of source regions have only one
massive cluster projected into them. This result is
clearly correlated directly with beam size, but also depends weakly on
sensitivity, since a deeper survey will lead to larger source regions
(when limited by signal-to-noise) and will increase the chance of
objects overlapping into the source, and can also increase
``bridging'' between sources. Also, it is likely that SZE experiments
will sometimes detect a source where there is no cluster in the mass
range at that location (spurious detections). Therefore our estimates
shown in Figure \ref{all_source} may be optimistic. 
 
\subsection{Contribution of Unresolved Halos and Unbound Gas}\label{sec:flux_break}

One possibly significant difference between this and other similar
simulation-based studies is the inclusion of adiabatic hydrodynamics
in addition to N-body dynamics in the calculation. This results in
several advantages over N-body calculations that include the effects
of baryons in the post-processing phase. The first is that our clusters
need not be in hydrostatic equilibrium (HSE) which is a standard
assumption in dark matter-only simulations which have been post-processed. Both simulations and
observations indicate that hydrostatic equilibrium is not a safe
assumption for many clusters \citep[see e.g.,][]{rasia,1E06}. A significant amount of scatter in cluster
observables results from the disequilibrium caused by mergers
\citep{roettmass, ricker, randall}. This
scatter is absent in N-body + HSE type studies, and is naturally
included in our work. Secondly, our simulations include baryons
which are outside virialized objects, including gas in filaments and voids. The
SZE in particular is sensitive to this additional gas, since the
effect is only linear in the gas density and is relatively redshift
independent. Thus any gas along the line of sight contributes to the
SZE integral, and is not diminished by distance. 

\begin{figure*}[t]
\begin{center}
\plottwo{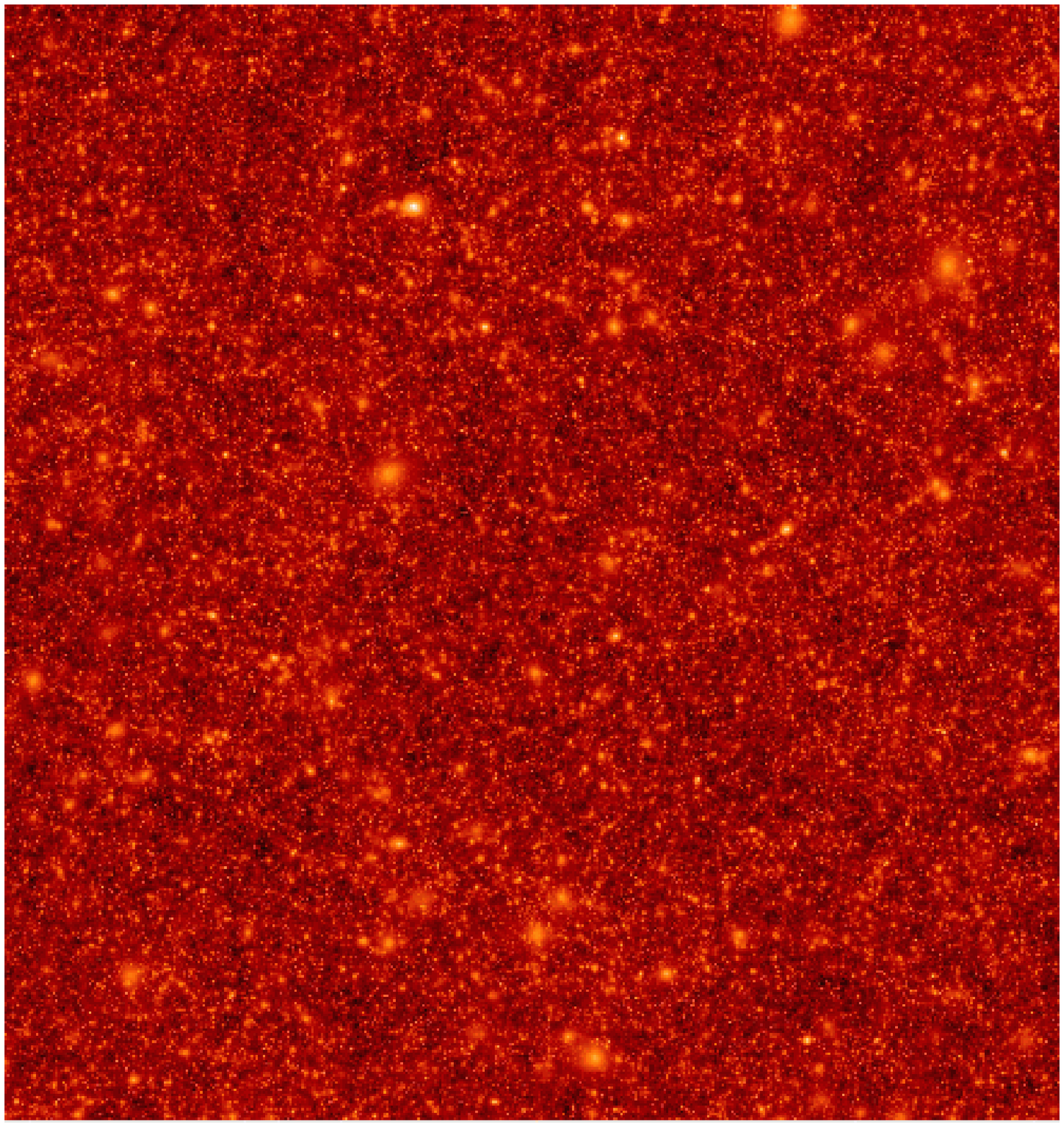}{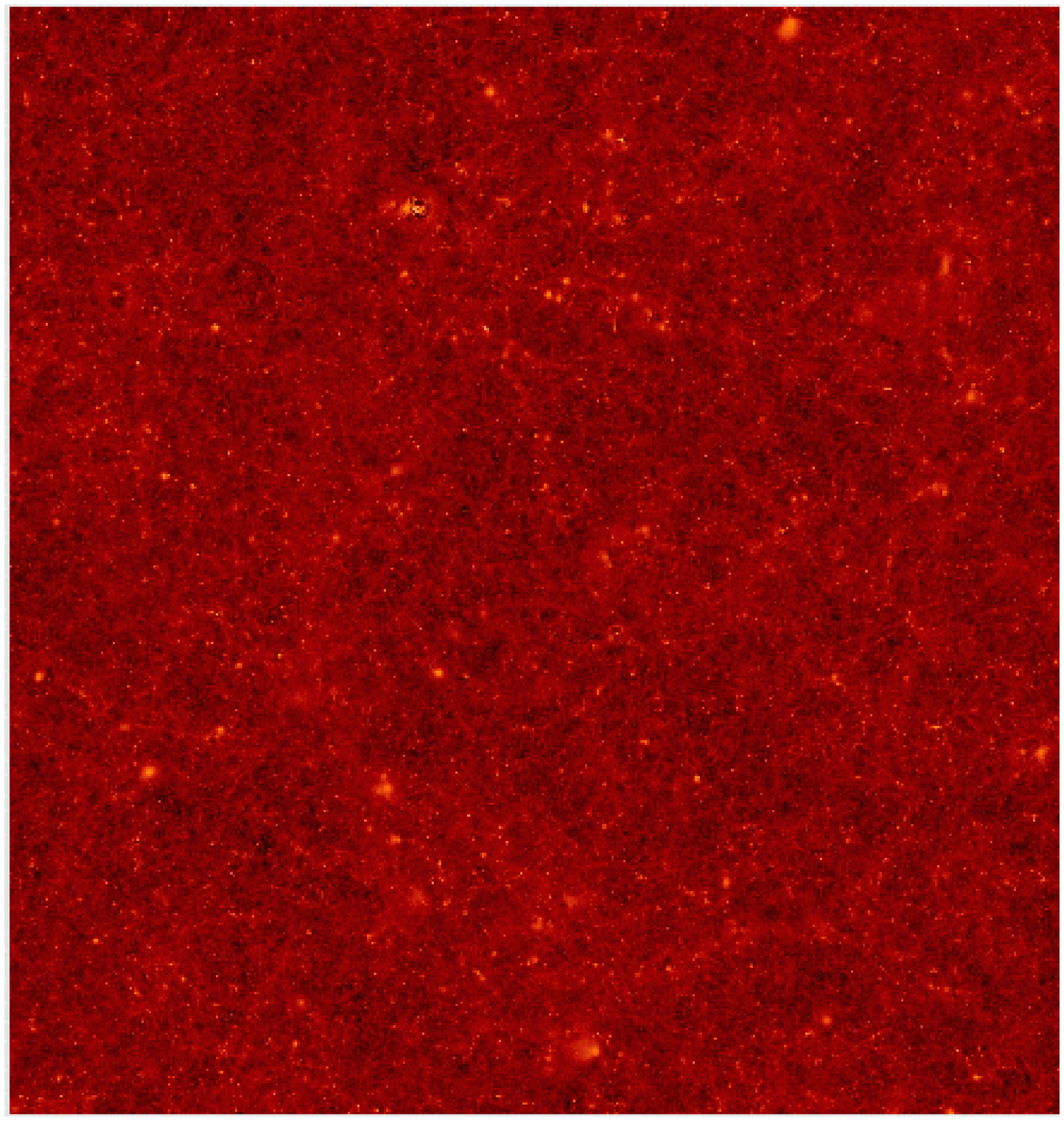}
\end{center}
\caption{Left panel: 100 deg$^2$ projected light cone image of the
  Compton y-parameter from a $512^3$ \textit{Enzo} simulation of a
  $(512h^{-1}$ Mpc$)^3$ volume with 7 dynamic levels of refinement. Light
  cone includes tiles at 27 discrete redshift intervals between z=3 and z=0.1. Right
panel: Same image as left panel, but with clusters with M $\geq 5 \times
  10^{13} M_{\odot}$ cut from the data. Roughly one third of the total flux in
  the image comes from the objects that remain after the removal of
  the massive clusters, including poor groups and filaments. We
  predict that such observations could provide the first detection of
  the WHIM gas over large sky areas.}
\label{sz_lc}
\end{figure*}
\begin{figure}
\begin{center}
\includegraphics[width=0.5\textwidth]{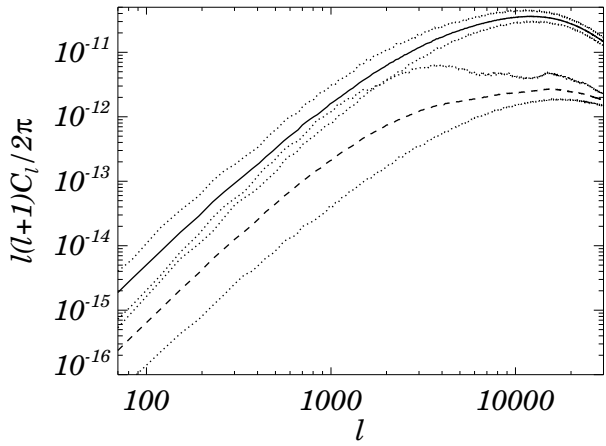}
\end{center}
\caption{Angular power from 200 full light cone images (solid line) compared to
  angular power from images where M $\geq 5 \times
  10^{13}M_{\odot}$ halos are subtracted (dashed line).  Dotted lines
  indicate 90\% variance range for the 200 independent stacking
  realizations of the light cone images.}
\label{power_noc}
\end{figure}

 
While several authors have noted that the angular power in the SZE
from the cluster subtracted field is small \citep[see,
e.g.,][]{holder}, it is not necessarily true that the total SZE flux
(or decrement) from unresolved halos and unbound gas is negligible.  Figure \ref{sz_lc} shows an image of the full field
of light cone in projected Compton $y$ parameter next to an
image of the field where clusters with M $> 5.0\times 10^{13}M_{\odot}$ have
been removed from the field. We
show in Figure \ref{power_noc} the angular power from the SZE in the
cluster subtracted images compared to the power due to the full SZE
image. At small scales ($\ell >$ 5000), the power in the cluster
subtracted image is more than an order of magnitude lower than that of
the full image. At larger scales, however, the difference narrows, and
in fact the 90\% variance overlaps for the two in some regions. This
is likely due partly to incomplete cluster subtraction in the image.
 
\begin{figure}
\begin{center}
\includegraphics[width=0.45\textwidth]{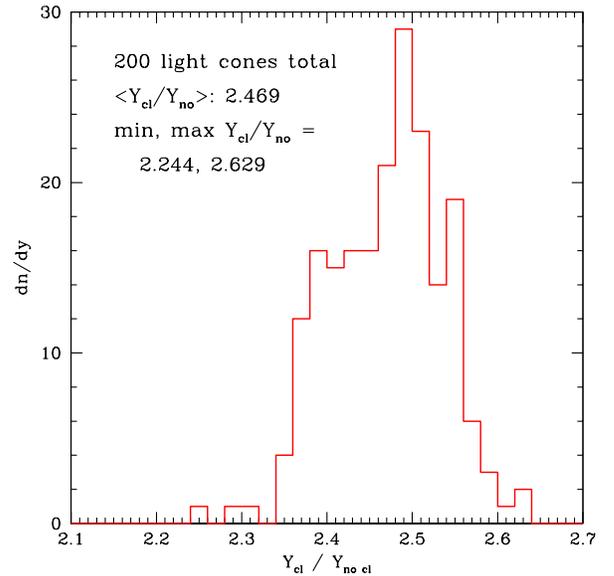}
\end{center}
\caption{Histogram of the ratio of total flux in the SZE y
    parameter images from clusters with M $\geq 5 \times
  10^{13}M_{\odot}$ and from images with clusters subtracted. The histogram is
    generated from 200 independent realizations of the
    light cone using the same simulation. Roughly two thirds of the flux in the
    image comes from clusters with M$\geq 5 \times
  10^{13}M_{\odot}$, and the other third comes from the WHIM and poor groups.}
\label{szy_histo}
\end{figure}

A significant result from this analysis is that
roughly one third of the SZE flux in the image comes from objects with M
  $< 5.0\times 10^{13}M_{\odot}$ and filamentary structures made up of
  gas in the Warm-Hot Intergalactic Medium (WHIM) phase. Figure \ref{szy_histo} shows a histogram for 200 independent light cone
realizations from our simulations of the ratio of the total SZE flux
(or integrated Compton $y$ parameter) in the 100 deg$^2$ field from
only gas within the virial radius of those clusters to that of all other gas
in the field. Thus, we predict that upcoming
SZE instruments are the only near-term telescopes that will possibly be capable
of detecting WHIM over large sky areas.  This result is consistent
with \citet{verde}, who have performed a similar study with a fixed
grid N-body/hydro simulation with considerably lower peak resolution
(195 h$^{-1}$ kpc) than our work.

It is important to note that in an adiabatic simulation, gas fractions
are relatively constant with cluster/group mass. In the real universe,
as well as in more realistic simulations such as those we performed
for \citet{hall06}, the ICM gas radiatively cools and forms stars,
lowering the gas fraction of the cluster, and effectively attenuating
its SZE signal. This effect is also cluster-mass dependent, lower mass
objects have comparatively lower gas fractions. Within a
simulation with radiative cooling and star formation prescriptions, we find gas
fractions 30-50\% lower in clusters with M $\approx 10^{14} M_{\odot}$
than the average value for clusters with M $> 3\times 10^{14}
M_{\odot}$. Additionally, recent observational studies
\citep[e.g.,][]{mcc07} point out that
gas fractions deduced from X-ray data decrease with cluster temperature (therefore with mass),
by of order 50\% for 1-2 keV clusters from a flat value above $\sim$4
keV.  Therefore we expect our result here to be an upper limit
on the amount of flux from low mass objects and WHIM gas, possibly
above the true value by as much as a factor of two. Our future light
cone simulations will be run with non-gravitational physics, and we
will explore the effect on the SZE background. 
 
\subsection{Contribution per Source of Cluster Subtracted Images}
Since unresolved halos (M $< 5 \times
  10^{13} M_{\odot}$) and unbound gas in this simulation clearly contribute flux to
  the image, it is important to ask how much additional
  flux per source is added.  This extra flux is a bias, in that
  it is always additive. Therefore it should boost the emission of all
  clusters in the field by some amount which may vary from cluster to
  cluster, adding both bias and scatter to the cluster SZE
  observable. This effect is critical to understand, since photometric
  accuracy of the SZE in clusters is key to calibrating a Y-M
  relationship which will be useful in determining cosmological paramteters. The precision of the calibration
  of the Y-M relationship depends strongly on its scatter
  \citep{melin}. 

\begin{figure}
\begin{center}
\epsscale{.9}
\includegraphics[width=0.5\textwidth]{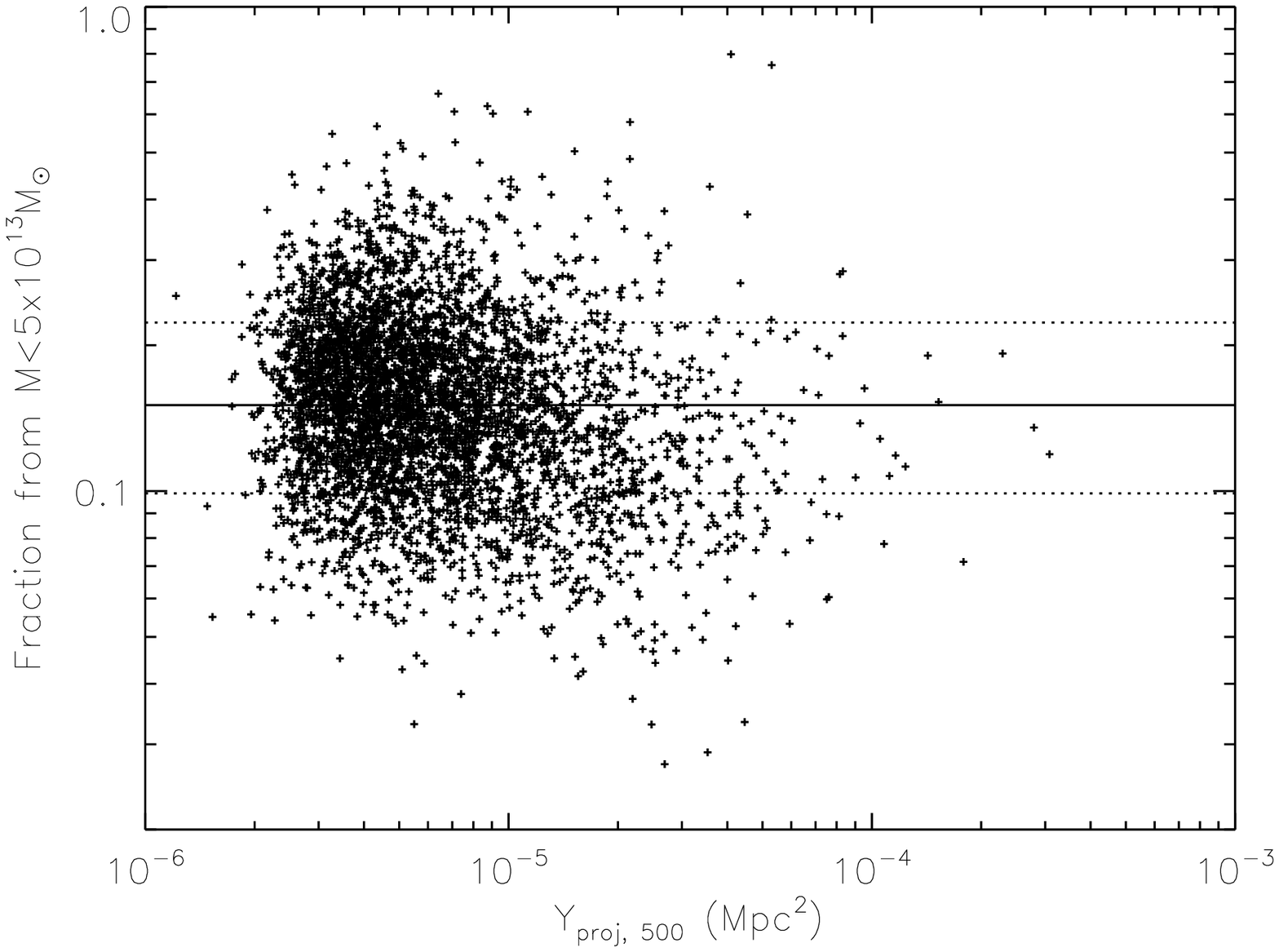}
\end{center}
\caption{Ratio of integrated Compton $y$ inside a cylinder projected to $r_{500}$ from the
  cluster-subtracted image to integrated Compton $y$ from the
same projected cylinder in the full light cone image. Includes all
  clusters projected into the survey image which have M $\geq 10^{14}
  M_{\odot}$. }
\label{whim_ratio}
\end{figure}

  Figure \ref{whim_ratio} shows the ratio of the integrated Compton $y$
  parameter inside a cylinder of radius $r_{500}$ around each cluster
  above 10$^{14} M_{\odot}$ from the
  cluster subtracted image to the value for the full image. The plot
  shows values for clusters in a single light cone image, but there is
  very little variation in the mean, median and scatter across all the
  realizations. The large scatter is partly a result of
  incomplete subtraction of the cluster SZE signature from the image,
  but also results from variations in large scale structure in the
  various source regions. What is clear from the plots is that there
  is a systematic bias in integrated Compton $y$ resulting from low
  mass objects and unbound gas. The mean value of this ratio is $16.3^{+7.0}_{-6.4}$\%. The
  1$\sigma$ scatter is $\pm$30-40\%. It is unclear
  that this bias is reducible, since identifying objects of very low
  mass in source fields, particularly at high redshift, seems
  prohibitively difficult. Additionally, making direct observation of
  filamentary gas, particularly to locate its position on the sky, has
  been nearly impossible. 
 
There will be some variation from survey to survey in this
additional flux. Depending on how much of the cluster's radial extent
is sampled, the mean value for all clusters will change. The scatter though,
is quite large, which means that accounting for this flux is not as
simple as removing a uniform background from each cluster's SZE
signal. In our study we also performed the identical analysis for the
clusters, assuming detection out to $r_{2500}$, and found no change in
the scatter, though the mean value of the additional flux dropped to
roughly 8-10\%. As discussed in previous sections, a more realistic
modeling of the heating and cooling in the ICM should result in a
reduction in this additional flux (or decrement) by as much as
30-50\%. Even with a reduction of this size, we still expect it to be
a few percent to 10\% effect with of order $\pm$30-40\% scatter, creating
challenges for a percent-level calibration of the Y-M relationship.

\section{Discussion and Summary}\label{sec:discuss}
In this study, we have taken an important step missing from previous work in the
literature on characterizing selection functions of SZE surveys. In
earlier work, investigators have attempted to evaluate methods of
removing contamination of the galaxy cluster SZE signal due to the CMB
and other point sources and backgrounds. However, they have included
only the baryons present in clusters and groups, and artificially
inserted the gas in hydrostatic equilibrium with the dark matter
distribution from an N-body only simulation. Here, we have explored
the often neglected contribution of gas in low mass halos and unbound
filamentary gas in aggregate to determine the effect on the cluster
signal at a single frequency. 

The presence of gas outside the cluster virial radii in low density
structures such as filaments is potentially an important
contributor to the cluster SZE signal. This gas is completely absent in
non-hydrodynamic treatments, but appears
naturally in our calculation. Since the SZE does not diminish with
distance and results from a line of sight integral of the gas
pressure, the sum of all the flux from unbound gas could be a
significant contributor to the total flux in any cluster
detection. Additionally, our simulation includes
the cluster gas in the full array of dynamically active states. Though
the integrated SZE is less sensitive to cluster-cluster merging than an X-ray
observation would be, merging contributes non-zero scatter to the Y-M
relationship. Thus, for all these reasons it is critically important to self-consistently include
baryons in numerical simulations in order to properly simulate sky surveys.  

We have shown that on 100 square degree patches of the sky
at $l < 2000$ or so the deduced power can be different by factors of
5-8. This indicates that the power spectrum can be quite different
from one area of the sky to another, and clearly requires much larger
angular areas to be well constrained. The effect of cosmic variance does
not become very small until $\ell > $ a few thousand. 

We have shown that projection effects can create a large bias and
additional scatter in the value of Y measured for clusters of
galaxies. For clusters of M $\approx$3.0$\times$10$^{14}
M_{\odot}$, the median bias in Y due to projection is 79\%, and the
scatter is +32\%/-16\% about that median, a significant increase over
the scatter in the true Y-M relationship. Additionally, we have shown
that the contribution of low-mass unresolved
halos and unbound gas to the flux (or decrement) of identified sources can be
significant in some cases, and certainly varies widely from source to
source. We find that there
is a contribution from gas outside clusters of $16.3^{+7.0}_{-6.4}$\% per object
on average for upcoming surveys. This indicates both a bias and an additional source of scatter
in the determination of the true SZE signal from any given cluster. As identified by other studies,
errors in extracting the correct value of Y should dominate the error
budget for the Y-M relation. This effect is critical to understand,
since photometric accuracy of the SZE in clusters is key to calibrating a Y-M
relationship which can be useful for the precision determination of
cosmological parameters. While the intrinsic Y-M relation
has very small scatter, what matters in practice is the ability to
determine the value of Y accurately. The precision of the calibration
of the Y-M relationship depends strongly on its scatter. 

We also show results from an analysis of the source confusion for each
instrument based on how many massive (M $>10^{14}$ M$_{\odot}$)
clusters lie within an identified source region (within a radius of
$r_{500}$ or a beam diameter at 144 GHz). It
appears that pure cluster/group confusion in these surveys will be a
significant problem, particularly for Planck Surveyor, but also to a
lesser extent for the other single-dish surveys. Smaller
beam surveys ($\sim$1$\arcmin$) have more than one massive
cluster within a beam diameter 5-10\% of the time, and a larger beam
experiment like Planck has multiple clusters per beam 60\% of the
time. We may have slightly
overestimated the problem, since the use of higher resolution (shorter
wavelength) bands on some of the survey instruments will help to
alleviate this issue. On the other hand, we have not accounted for
spurious detections which may result in a field with real backgrounds
and instrument noise. Whether this type of confusion can be mitigated
depends on a variety of factors, including the mass of the additional
secondary objects in the source region, the redshift distribution and
angular scale of the objects and the observing beam, and multiple
wavelength identfication of the objects (e.g., using optical, X-ray,
and lensing measurements).

This study uses a large volume, high resolution adiabatic
simulation, which serves as a template for more complex runs involving
additional non-gravitational physics that is likely important to
accurately modeling cluster surveys. There remain important
differences between simulation outputs and cluster observations, particularly
for lower mass clusters, which indicates the need for a better
understanding of the details of baryonic cluster physics. In addition, deviations from isothermality and
hydrostatic equilibrium in the cluster gas can have a strong impact on
both the observable and derived properties of clusters. 
It is also important to note that there is some dependence of the SZE signal
on the details of the ICM physics (heating, cooling, conduction, etc.)
which is not modeled in this work. In future work, we will explore the
impact of this additional physics, modeled self-consistently within
the hydrodynamic framework of the simulation code, on a selection of SZE
clusters from surveys. 

This work will be expanded in future papers by a detailed treatment of
the point source confusion and instrumental and observing
limitations. This will include adding to our synthetic surveys the
contribution of the CMB, dusty galaxies, AGN, and atmospheric
foreground. We are now working on modeling these, in particular for
APEX-SZ and SPT, and will then experiment with matched filtering and
the use of multiwavelength coverage provided by SPT for example to
mitigate confusion and remove the atmospheric and CMB signals. Matched
filtering is a process by which the images are filtered with a kernel
matched to the presumed size and shape of the expected sources in the
field. In the case of the SZE, this procedure filters out information
on larger scales where the primary CMB anisotropies will be a source
of confusion. It also maximizes the contrast of the image for the
objects at that scale. Since the angular scale subtended by massive
clusters is a function of redshift and a weak function of cluster
mass, spatial filtering will need to be done at a variety of angular
scales to get a complete cluster sample as in \citet{melin}. A similar type
of analysis has been performed by \citet{sehgal} for ACT's survey. 

We also are currently performing additional synthetic light cone sky
surveys at X-ray wavelengths, to take a first look at the limitations
of current (e.g. XMM-LSS) and upcoming (e.g. eRosita) X-ray surveys in
extracting cosmological parameters. 
\acknowledgments
BWO and MLN have been supported in part by NASA
grant NAG5-12140 and NSF grant AST-0307690. 
BWO has been funded in part
under the auspices of the U.S.\ Dept.\ of Energy, and supported by its
contract W-7405-ENG-36 to Los Alamos National Laboratory.  The
simulations were by performed at SDSC and NCSA with computing time provided by 
NRAC allocation MCA98N020. EJH and JOB have been supported in part by a grant 
from the U.S. National Science Foundation (AST-0407368). EJH also
acknowledges support from NSF AAPF AST-0702923. We thank Yoel Rephaeli for useful discussions on the
SZE angular power spectrum. We thank John Carlstrom, Nils Halverson, Jeremiah Ostriker
,Douglas Rudd and an anonymous referee for useful comments and critiques. We thank the Aspen Center
for Physics for hosting three of the authors during the writing of this manuscript,
and also thank Maria's New Mexican Kitchen in Santa Fe for providing an atmosphere conducive
to the conception of this project.

\bibliographystyle{apj}

\end{document}